\documentclass[letterpaper]{article} 
\usepackage{aaai24}  
\usepackage{times}  
\usepackage{helvet}  
\usepackage{courier}  
\usepackage[hyphens]{url}  
\usepackage{graphicx} 
\usepackage{natbib}  
\usepackage{caption} 
\usepackage{algorithm}
\usepackage{algorithmic}

\usepackage{newfloat}
\usepackage{listings}
\DeclareCaptionStyle{ruled}{labelfont=normalfont,labelsep=colon,strut=off} 
\floatstyle{ruled}
\newfloat{listing}{tb}{lst}{}
\floatname{listing}{Listing}
\usepackage{tikz}
\usepackage{amssymb}
\usepackage{amsthm}
\usepackage{amsmath}
\usepackage{amsfonts}
\usepackage{mathabx}
\usepackage{mathtools}
\usepackage{listings}
\usepackage{caption}
\usepackage{float}
\usepackage{natbib}
\usepackage{tcolorbox}
\usepackage{subcaption}
\usetikzlibrary{patterns,automata,positioning,arrows}
\usepackage{bm}

\usepackage{pgfplots}
\usepgfplotslibrary{statistics} 
\usepgfplotslibrary[statistics] 
\usetikzlibrary{pgfplots.statistics} 
\usetikzlibrary[pgfplots.statistics] 
\pgfplotsset{width=6cm,compat=1.18}\usepgfplotslibrary{statistics}

\title{Data Poisoning to Fake a Nash Equilibrium in Markov Games}

\author{
    Young Wu,
    Jeremy McMahan,
    Xiaojin Zhu,
    Qiaomin Xie
}
\affiliations{
    University of Wisconsin - Madison \\
    yw@cs.wisc.edu, jmcmahan@wisc.edu, jerryzhu@cs.wisc.edu, qiaomin.xie@wisc.edu

}

\usepackage{bibentry}

\global\long\def\Ddag{D^{\dagger}}

\global\long\def\Nh{N_{h}}
\global\long\def\Ph{P_{h}}
\global\long\def\Phh{\hat{P}_{h}}

\global\long\def\QHH{Q_{H+1}}
\global\long\def\Qh{Q_{h}}

\global\long\def\Qhh{Q_{h+1}}

\global\long\def\Qhhh{\hat{Q}_{h}}

\global\long\def\Qlh{\underline{Q}_{h}}

\global\long\def\Qlhh{\underline{Q}_{h+1}}

\global\long\def\Quh{\overline{Q}_{h}}

\global\long\def\Quhh{\overline{Q}_{h+1}}

\global\long\def\Rh{R_{h}}

\global\long\def\Rhh{\hat{R}_{h}}

\global\long\def\TOM{\mathcal{T}}

\global\long\def\ai{a_{i}}
\global\long\def\ak{\textbf{a}^{(k)}}
\global\long\def\akh{\textbf{a}^{(k)}_{h}}

\global\long\def\bMPE{\underline{\mathcal{U}}}
\global\long\def\bNash{\underline{\mathcal{U}}}

\global\long\def\bTOM{\overline{\mathcal{T}}}

\global\long\def\fMPE{\mathcal{M}}
\global\long\def\fNash{\mathcal{N}}

\global\long\def\iMPE{\mathcal{U}}
\global\long\def\iNash{\mathcal{U}}

\global\long\def\mA{\mathcal{A}}
\global\long\def\mAa{\mathcal{A}_{1}}
\global\long\def\mAb{\mathcal{A}_{2}}
\global\long\def\mAi{\mathcal{A}_{i}}

\global\long\def\mCPh{\mathcal{C}^{(P)}_{h}}

\global\long\def\mCRh{\mathcal{C}^{(R)}_{h}}

\global\long\def\mD{\mathcal{D}}

\global\long\def\mQ{\mathcal{Q}}
\global\long\def\mR{\mathcal{R}}
\global\long\def\mS{\mathcal{S}}

\global\long\def\pia{\pi_{1}}
\global\long\def\pib{\pi_{2}}
\global\long\def\pidag{\pi^{\dagger}}
\global\long\def\pidagh{\pi^{\dagger}_{h}}

\global\long\def\pidagha{\pi^{\dagger}_{1,h}}
\global\long\def\pidaghb{\pi^{\dagger}_{2,h}}

\global\long\def\pih{\pi_{h}}
\global\long\def\piha{\pi_{1,h}}
\global\long\def\pihb{\pi_{2,h}}

\global\long\def\pii{\pi_{i}}
\global\long\def\piih{\pi_{i,h}}

\global\long\def\rdag{r^{\dagger}}
\global\long\def\rdagk{r^{\dagger,(k)}}
\global\long\def\rdagkh{r^{\dagger,(k)}_{h}}
\global\long\def\rhoQ{\rho^{(Q)}}
\global\long\def\rhoQh{\rho^{(Q)}_{h}}
\global\long\def\rhoP{\rho^{(P)}}
\global\long\def\rhoPh{\rho^{(P)}_{h}}
\global\long\def\rhoR{\rho^{(R)}}
\global\long\def\rhoRh{\rho^{(R)}_{h}}

\global\long\def\rk{r^{(k)}}
\global\long\def\sk{s^{(k)}}
\global\long\def\rkh{r^{(k)}_{h}}

\global\long\def\skh{s^{(k)}_{h}}
\global\long\def\skhh{s^{(k)}_{h+1}}

\begin{document}

\newtheorem{thm}{Theorem}
\newtheorem{cor}{Corollary}
\newtheorem{lem}{Lemma}
\newtheorem{prop}{Proposition}
\newtheorem{conj}{Conjecture}
\newtheorem{algo}{Algorithm}
\newtheorem{obs}{Observation}
\newtheorem{clm}{Claim}
\newtheorem{df}{Definition}
\newtheorem{eg}{Example}
\newtheorem{asm}{Assumption}
\newtheorem{cond}{Condition}
\newtheorem{rmk}{Remark}

\newcommand{\jerry}[1]{\textcolor{red}{[jerry: #1]}}
\newcommand{\jeremy}[1]{\textcolor{orange}{[jeremy: #1]}}
\newcommand{\yudong}[1]{\textcolor{blue}{[yudong: #1]}}
\newcommand{\qiaomin}[1]{\textcolor{magenta}{[qiaomin: #1]}}
\newcommand{\young}[1]{\textcolor{green}{[young: #1]}}

\allowdisplaybreaks
\maketitle

\begin{abstract}
We characterize offline data poisoning attacks on Multi-Agent Reinforcement Learning (MARL), where an attacker may change a data set in an attempt to install a (potentially fictitious) unique Markov-perfect Nash equilibrium for a two-player zero-sum Markov game. We propose the unique Nash set, namely the set of games, specified by their Q functions, with a specific joint policy being the unique Nash equilibrium. The unique Nash set is central to poisoning attacks because the attack is successful if and only if data poisoning pushes all plausible games inside the set. The unique Nash set generalizes the reward polytope commonly used in inverse reinforcement learning to MARL. For zero-sum Markov games, both the inverse Nash set and the set of plausible games induced by data are polytopes in the Q function space. We exhibit a linear program to efficiently compute the optimal poisoning attack. Our work sheds light on the structure of data poisoning attacks on offline MARL, a necessary step before one can design more robust MARL algorithms.
\end{abstract}

\section{Introduction} 
Data poisoning attacks have been well studied in supervised learning (intentionally forcing the learner to train a wrong classifier) and reinforcement learning (wrong policy)~\cite{banihashem2022admissible,  huang2019deceptive, liu2021provably, rakhsha2021policy, rakhsha2021reward, rakhsha2020policy, sun2020vulnerability, zhang2020adaptive, ma2019policy, rangiunderstanding, zhang2008value, zhang2009policy}. 
Can data poisoning attacks be a threat to Markov Games, too?
This paper answers this question in the affirmative:
Under mild conditions, an attacker can force two game-playing agents to adopt any fictitious Nash Equilibrium (NE), which does not need to be a true NE of the original Markov Game.
Furthermore, the attacker can achieve this goal while minimizing its attack cost, which we define below.
Clearly, such power poses a threat to the security of Multi-Agent Reinforcement Learning (MARL).

Formally, we study two-player zero-sum Markov game offline data poisoning, stated as the following. \smallskip

\noindent\textbf{Problem Statement: Offline Data Poisoning.}
Let $D$ be a dataset $\{(\sk, \ak, \rk)\}_{k=1}^{K}$ with $K$ tuples of state $s$, joint action $\textbf{a} = (a_1, a_2)$, rewards $(r, -r)$.
The attacker's target NE is an arbitrary pure strategy pair $\pi^\dagger := (\pi^\dagger_1, \pi^\dagger_2)$.
The attacker can poison $D$ into another dataset $D^\dagger$ by paying cost $C(D,D^\dagger)$.
Two MARL agents then receive $D^\dagger$ instead of $D$.
The attacker aims to enforce that the agents learn the target NE $\pi^\dagger$ from $D^\dagger$ while minimizing $C$.

This problem is not well studied in the literature. Naive approaches -- such as modifying all the actions in the dataset to those specified by the target policy $(\pi^\dagger_1, \pi^\dagger_2)$ -- might not achieve the attack goal for MARL learners who assign penalties due to the lack of data coverage. 
Modifying all the rewards in the dataset that coincide with the target policy to the reward upper bound might be feasible, but would not be optimal in terms of attack cost $C$. 
Results on data poisoning against single-agent RL cannot be directly applied to the multi-agent case. In particular, there are no optimal policies in MARL, and equilibrium policies are computed instead. There could be multiple equilibria that are significantly different, and consequently, installing a target policy as the unique equilibrium is difficult. To resolve this issue, we provide a novel characterization of when a zero-sum Markov game has a unique Markov perfect Nash equilibrium.

Our framework can be summarized by the mnemonic ``ToM moves to the UN''.   
(i) UN stands for the Unique Nash set, which is the set of Q functions that make the target $\pi^\dagger$ the unique NE.
Uniqueness is crucial for the attacker to ensure that MARL agents choose the target NE with certainty, without breaking ties arbitrarily among multiple NEs.
(ii) ToM stands for the attacker's Theory of Mind of the MARL agents, namely 
the plausible set of Q functions 
that the attacker believes 
the agents will entertain upon receiving the poisoned dataset $D^\dagger$.
(iii) The attack is successful if, by controlling $D^\dagger$, the ToM set is moved inside the UN set.
A successful attack with the smallest cost $C(D,D^\dagger)$ is optimal.

Adversarial attacks on MARL have been studied in some recent work~\cite{ma2021game, gleave2019adversarial, guo2021adversarial}, but we are only aware of one previous work~\cite{wu2022reward} on offline reward poisoning against MARL.
Nonetheless, they require a strong assumption of full data coverage, and that the learners compute the Dominant Strategy Markov Perfect Equilibrium (DSMPE).
In contrast, we do not require full coverage, and we consider a weaker solution concept, Markov Perfect Equilibrium (MPE).
Our general attack framework also accommodates other forms of data poisoning.

Understanding adversarial attacks in the multi-agent setting is critical since many real-life applications of MARL problems are susceptible to adversarial attacks. Examples of two-player zero-sum games include board games such as GO and Chess~\cite{silver2017mastering, silver2016mastering}, where the learners use historical game plays as training data and an attacker can potentially alter the data to change the behavior of the trained agents. In the case of competitive robotics, for example, robot soccer~\cite{gu2017deep, riedmiller2009reinforcement, kober2013reinforcement}, they are trained on offline datasets and the attacker can mislead the trained policies by modifying the training sets. For finance application, especially algorithmic or high-frequency stock or option trading~\cite{lee2007multiagent, lee2002multi} that are usually trained on historical prices, if the database is corrupted by an attacker, the learned trading strategies can be sub-optimal as well. There are also examples of multi-player games that have two-player games as special cases, for example, video games~\cite{vinyals2019grandmaster, jaderberg2019human, berner2019dota}, card games~\cite{brown2019superhuman, brown2017libratus}, autonomous driving~\cite{shalev2016safe}, automated warehouses~\cite{yang2020multi}, and economic policymaking, which can all be trained on offline datasets and become vulnerable to adversarial attacks. In all of the above MARL applications, the threat of adversarial attacks has not been investigated.




Our contributions include a unified framework for offline data poisoning attacks, and in particular, a linear program formulation that efficiently solves the reward poisoning problem for two-player zero-sum Markov games. On the technical side, we present a geometric characterization of a deterministic policy being the unique Markov perfect Nash equilibrium of zero-sum Markov games. In addition, we demonstrate that for a class of MARL learners that compute equilibrium policies based on games within confidence regions around a point estimate of the Q function of the Markov game, an attack with appropriate parameters on these learners would success on most of the model-based and model-free offline MARL learners proposed in the literature.

\section{Offline Attack on a Normal-form Game} 

\subsection{The Unique Nash Set (UN) of a Normal-form Game}
We present the main components of our approach with a normal-form game, in particular, a two-player zero-sum game is a tuple $\left(\mA, R\right)$, where $\mA = \mAa \times \mAb$ is the joint action space and $R : \mA \to  \left[-b, b\right]$ is the mean reward function. We use $b = \infty$ in the case of unbounded rewards. Given $\mA$, we denote the set of reward functions by $\mR = \left\{ R : \mA \to  \mathbb{R} \right\}$.

A pure strategy profile $\pi = \left(\pia, \pib\right)$ is a pair of actions, where $\pii \in \mAi$ specifies the action for agent $i \in \left\{1, 2\right\}$. We focus on pure strategies, but we allow mixed strategies in which case we use the notation $\pii\left(\ai\right)$ to represent the probability of $i$ using the action $\ai \in \mAi$, and $R$ computes the expected reward $R\left(\pi\right) \coloneqq \displaystyle\sum_{a_{1} \in \mAa, a_{2} \in \mAb} \pia\left(a_{1}\right) \pib\left(a_{2}\right) R\left(\left(a_{1}, a_{2}\right)\right)$.

\begin{df} [Nash Equilibrium] \label{df:ne} 
A Nash equilibrium (NE) of a normal-form game $\left(\mA, R\right)$ is a mixed strategy profile $\pi$ that satisfies,
\begin{align}
R\left(\left(\pia, a_{2}\right)\right) &= R\left(\pi\right) = R\left(\left(a_{1}, \pib\right)\right), \nonumber
\\ & \forall\; a_{1} : \pia\left(a_{1}\right) > 0, a_{2} : \pib\left(a_{2}\right) > 0,\nonumber
\\ R\left(\left(\pia, a_{2}\right)\right) &\leq R\left(\pi\right) \leq R\left(\left(a_{1}, \pib\right)\right), \nonumber
\\ & \forall\; a_{1} : \pib\left(a_{1}\right) = 0, a_{2} : \pib\left(a_{1}\right) = 0,\nonumber
\end{align}
in particular, for a pure strategy profile $\pi$, it is a Nash equilibrium if,
\begin{align}
R\left(\left(\pia, a_{2}\right)\right) &\leq R\left(\pi\right) \leq R\left(\left(a_{1}, \pib\right)\right), \label{eq:ene}
\\ & \forall\; a_{1} \neq \pia, a_{2} \neq \pib. \nonumber 
\end{align}
We define $\fNash\left(R\right) \coloneqq \left\{\pi : \pi \text{\;is an NE of\;} \left(\mA, R\right) \right\}$ to be the set of all Nash equilibria of a normal-form game $\left(\mA, R\right)$.

\end{df}
Now, we define the inverse image of $\fNash$ from a single pure strategy profile $\pi$ back to the space of reward functions to be the unique Nash set.

\begin{df} [Unique Nash] \label{df:un} 
The unique Nash set of a pure strategy profile $\pi$ is the set of reward functions $R$ such that $\left(\mA, R\right)$ has a unique Nash equilibrium $\pi$,
\begin{align}
\iNash\left(\pi\right) &\coloneqq \fNash^{-1}\left(\left\{\pi\right\}\right) = \left\{ R \in \mR : \fNash\left(R\right) = \left\{\pi\right\} \right\}. \label{eq:eun}
\end{align}\end{df}
To characterize $\iNash\left(\pi\right)$, we note that for normal-form games, a pure strategy profile $\pi$ is the unique Nash equilibrium of a game if and only if it is a strict Nash equilibrium, which is defined as a policy $\pi$ that satisfies~\eqref{eq:ene} with strict inequalities.

\begin{prop} [Unique Nash Polytope] \label{prop:unp} 
For any pure strategy profile $\pi$,
\begin{align}
\iNash\left(\pi\right) &= \left\{ R \in \mR : \pi \text{\;is a strict NE of\;} \left(\mA, R\right) \right\} \nonumber
\\ &= \left\{ R \in \mR : R\left(\left(\pia, a_{2}\right)\right) < R\left(\pi\right) < R\left(\left(a_{1}, \pib\right)\right),\right. \nonumber
\\ & \hspace{2 em} \left. \forall\; a_{1} \neq \pia, a_{2} \neq \pib \right\}. \label{eq:eunp}
\end{align}\end{prop}
Here, the uniqueness is among all Nash equilibria including mixed-strategy Nash equilibria. The proof of the equivalence between~\eqref{eq:eun} and~\eqref{eq:eunp} is in the appendix. We restrict our attention to pure-strategy equilibria and defer the discussion of mixed strategy profiles to the last section.

To avoid working with strict inequalities, we define a closed subset of $\iNash\left(\pi\right)$ of reward functions that lead to strict Nash equilibria with an $\iota$ reward gap, which means all strict inequalities in~\eqref{eq:eunp} are satisfied with a gap of at least $\iota$, for some $\iota > 0$.

\begin{df} [Iota Strict Unique Nash] \label{df:isun} 
For $\iota > 0$, the $\iota$ strict unique Nash set of a pure strategy profile $\pi$ is, $\bNash\left(\pi; \iota\right) \coloneqq$
\begin{align}
& \left\{ R \in \mR : R\left(\left(\pia, a_{2}\right)\right) + \iota \leq R\left(\pi\right) \leq R\left(\left(a_{1}, \pib\right)\right) - \iota, \right. \nonumber
\\ & \hspace{2 em} \left. \forall\; a_{1} \neq \pia, a_{2} \neq \pib \right\}. \label{eq:eisun}
\end{align}\end{df}
For every pure strategy profile $\pi$ and $\iota > 0$, we have $\bNash\left(\pi; \iota\right) \subset \iNash\left(\pi\right)$, and the set is a polytope in $\mR$.

\subsection{The Attacker's Theory of Mind (ToM) for Offline Normal-form Game Learners}
We provide a model of the attacker's theory of mind of the victim, which is the attacker's belief about the learning algorithm the victim uses. In particular, the attacker is not required to have complete knowledge of the victims' learning algorithms: only an
approximation (of theory of mind) is needed. Formally, we define the theory-of-mind set as the set of plausible rewards that the victim uses based on the given training dataset, and we assume that the victims compute the Nash equilibria based on the reward functions estimated from a dataset $D \in \mD$, where $\mD$ is the set of possible datasets with $K$ episodes in the form $\left\{\left(\ak, \rk\right)\right\}_{k=1}^{K}$ , with $\ak \in \mA$ and $\rk \in \left[-b, b\right]$ for every $k \in \left[K\right]$.

\begin{df} [Theory of Mind] \label{df:tom} 
Given a dataset $D \in \mD$, the theory-of-mind set $\TOM\left(D\right) \subseteq \mR$ is the set of plausible reward functions that the victims estimate based on $D$ to compute their equilibria. In particular, if the victims learn an action profile $\pi$, then $\pi \in \displaystyle\bigcup_{R \in \TOM\left(D\right)} \fNash\left(R\right)$.

\end{df}
The theory-of-mind sets can be arbitrary and could be difficult to work with. We define an outer approximation the set that is a hypercube in $\mR$.

\begin{df} [Outer Approximation of Theory of Mind] \label{df:oatom} 
An outer approximation of $\TOM\left(D\right)$ is a set denoted by $\bTOM\left(D\right)$ that satisfies $\TOM\left(D\right) \subseteq \bTOM\left(D\right)$ for every $D \in \mD$, and can be written in the form, $\bTOM\left(D\right) \coloneqq$
\begin{align}
 \left\{ R \in \mR : \left| R\left(\textbf{a}\right) - \hat{R}\left(\textbf{a}\right) \right| \leq \rhoR\left(\textbf{a}\right), \forall\; \textbf{a} \in \mA \right\}, \label{eq:eoatom}
\end{align}
for some point estimate $\hat{R}$ and radius $\rhoR$.
\\* We call $\bTOM\left(D\right) \text{\;a\;}$ linear outer approximation if $\hat{R}$ is linear in $\left\{\rk\right\}_{k=1}^{K.}$

\end{df}
We present a few examples of the theory-of-mind sets as follows.

\begin{eg} [Theory of Mind for Maximum Likelihood Victims] \label{eg:tommle} 
Given a dataset $D \in \mD$, if the attacker believes the victims are maximum likelihood learners, then $\TOM\left(D\right)$ is a singleton $R^{\text{\;MLE\;}}$, where, for every $\textbf{a} \in \mA$,
\begin{align}
R^{\text{\;MLE\;}}\left(\textbf{a} | r\right) &\coloneqq \begin{cases} \dfrac{1}{N\left(\textbf{a}\right)} \displaystyle\sum_{k=1}^{K} \rk \mathbb{I}_{\left\{\ak = \textbf{a}\right\}} & \text{\;if\;} N\left(\textbf{a}\right) > 0 \\ 0 & \text{\;if\;} N\left(\textbf{a}\right) = 0 \\ \end{cases} \nonumber
\\ N\left(\textbf{a}\right) &\coloneqq \displaystyle\sum_{k=1}^{K} \mathbb{I}_{\left\{\ak = \textbf{a}\right\}}. \label{eq:emle}
\end{align}
The smallest outer approximation $\bTOM\left(D\right)$ can be specified using $\hat{R} = R^{\text{\;MLE\;}}$ and $\rhoR = 0$, and $\bTOM$ is linear since~\eqref{eq:emle} is linear in $\left\{\rk\right\}_{k=1}^{K}$ .

\end{eg}
\begin{eg} [Theory of Mind for Pessimistic Optimistic Victims] \label{eg:tompovi} 
Given a dataset $D \in \mD$, if the attacker believes the victims are learners that use pessimism and optimism by adding and subtracting bonus terms and estimating one or two games, as in~\cite{cui2022offline}, then $\TOM\left(D\right)$ may contain two reward functions $\underline{R}$ and $\overline{R}$, where for every $\textbf{a} \in \mA$,
\begin{align}
\underline{R}\left(\textbf{a}| r\right) &\coloneqq R^{\text{\;MLE\;}}\left(\textbf{a}| r\right) - \beta\left(\textbf{a}\right) \nonumber
\\ \overline{R}\left(\textbf{a}| r\right) &\coloneqq R^{\text{\;MLE\;}}\left(\textbf{a}| r\right) + \beta\left(\textbf{a}\right), \label{eq:epovi}
\end{align}
with $\beta\left(\textbf{a}\right) = \dfrac{c}{\sqrt{N\left(\textbf{a}\right)}}$ being the bonus term, for some constant $c$.
\\* The smallest outer approximation $\bTOM\left(D\right)$ can be specified using $\hat{R} = R^{\text{\;MLE\;}}$ and $\rhoR\left(\textbf{a}\right) = \beta\left(\textbf{a}\right)$ for every $\textbf{a} \in \mA$, and $\bTOM$ is linear since~\eqref{eq:emle} and~\eqref{eq:epovi} are both linear in $\left\{\rk\right\}_{k=1}^{K}$ .

\end{eg}
\begin{eg} [Theory of Mind for Data Splitting Victims] \label{eg:tomds} 
Given a dataset $D \in \mD$, if the attacker believes the victims use maximum likelihood estimates on a subsample of the $D$, similar to the data-splitting procedure in~\cite{cui2022offline}, then $\bTOM\left(D\right)$ could be viewed as a high-probability set of rewards that the victims are estimating and $\rhoR$ would be half of the confidence interval width for the mean of the subsample around the mean of the complete dataset $R^{\text{\;MLE\;}}$.

\end{eg}

\subsection{The Cheapest Way to Move ToM into UN for Normal-form Games}
The goal of the attacker is to install a specific action profile as the unique Nash equilibrium of the game learned by the victim while minimally modifying the training data. We consider a general attacker's cost as a function $C : \mD \times \mD \to  \mathbb{R}^{+}$ where $C\left(D, \Ddag\right)$ is the cost of modifying the dataset from $D$ to $\Ddag$. Given the original data set $D \in \mD$, the attacker's attack modality $\mD\left(D\right)$ is the set of datasets the attacker is allowed to modify the original dataset to. For the reward poisoning problem, where $\mD^{\left(R\right)}\left(D\right)$ is all possible datasets in which only rewards are modified from $\rk$ to $\rdagk$, we consider the following cost function.

\begin{eg} [$L_{1}$ Cost Function] \label{eg:lcf} 
For reward poisoning problems, we define the $L_{1}$ cost of modifying the dataset from $D = \left\{\left(\ak, \rk\right)\right\}_{k=1}^{K}$ to $\Ddag = \left\{\left(\ak, \rdagk\right)\right\}_{k=1}^{K}$ by $C^{\left(1\right)}\left(D, \Ddag\right) \coloneqq \displaystyle\sum_{k=1}^{K} \left| \rk - \rdagk \right|$.

\end{eg}

\begin{rmk} In our framework, the attacker’s cost function can be an arbitrary convex function, which can accommodate various settings, for example, when the attacker has a limited budget or when the attacker can only change a limited number of entries: the optimization will remain a convex program with linear constraints. $L_{1}$ loss is used for simplicity so that our attack optimization is a linear program, it could be relaxed, although then the optimization would be harder to solve.
\end{rmk}

Now, given the original dataset $D$ and the attacker's target action profile $\pidag$, we formally state the attacker's problem as finding the cheapest (minimal cost) way to move $\TOM\left(D\right)$ into $\iNash\left(\pidag\right)$.

\begin{df} [Attacker's Problem] \label{df:ap} 
The attacker's problem with the target action profile $\pidag$ is,
\begin{align}
\displaystyle\inf_{\Ddag \in \mD\left(D\right)} & C\left(D, \Ddag\right) \label{eq:eap}
\\ \;s.t.\; & \TOM\left(\Ddag\right) \subseteq \iNash\left(\pidag\right).\nonumber
\end{align}\end{df}
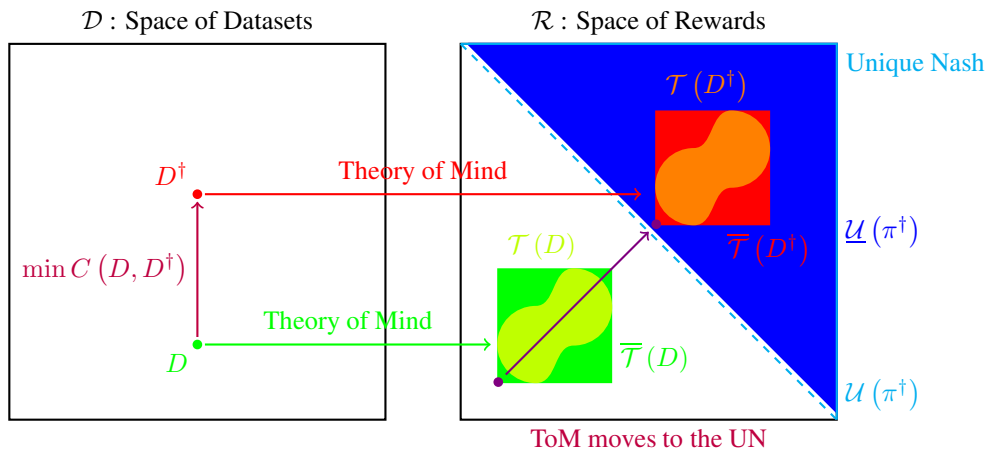
\begin{figure*}  \centering \begin{tikzpicture} [scale = 1] 
\draw[thick] (0.0, 0.0) rectangle (5.0, 5.0);
\node[above] at (2.5, 5.0){$\mD$ : Space of Datasets};
\draw[thick] (6.0, 0.0) rectangle (11.0, 5.0);
\node[above] at (8.5, 5.0){$\mR$ : Space of Rewards};
\draw[blue, fill = blue, thick] (6.1, 5.0) -- (11.0, 0.1) -- (11.0, 5.0) -- (6.1, 5.0);
\node[blue, right] at (11.0, 2.5){$\bNash\left(\pidag\right)$};
\draw[cyan, dashed, thick] (6.0, 5.0) -- (11.0, 0.0);
\node[cyan, below right] at (11.0, 5.0){Unique Nash};
\draw[cyan, thick] (6.0, 5.0) -- (11.0, 5.0) -- (11.0, 0.0);
\node[cyan, above right] at (11.0, 0.0){$\iNash\left(\pidag\right)$};
\draw[green, fill=green, thick] (2.5, 1.0) circle [radius = 0.05];
\node[green, below left] at (2.5, 1.0){$D$};
\draw[red, fill=red, thick] (2.5, 3.0) circle [radius = 0.05];
\node[red, above left] at (2.5, 3.0){$\Ddag$};
\draw[->, green, thick] (2.6, 1.0) -- (6.4, 1.0);
\node[green, above] at (4.5, 1.0){Theory of Mind};
\draw[->, red, thick] (2.6, 3.0) -- (8.4, 3.0);
\node[red, above] at (5.5, 3.0){Theory of Mind};
\draw[->, purple, thick] (2.5, 1.1) -- (2.5, 2.9);
\node[purple, left] at (2.5, 2.0){$\displaystyle\min C\left(D, \Ddag\right)$};
\draw[green, fill=green, thick] (6.5, 0.5) rectangle (8.0, 2.0);
\node[green, above right] at (8.0, 0.5){$\bTOM\left(D\right)$};
\draw[lime, fill=lime, thick] (6.5, 1.0)to [out = 270, in = 180](7.0, 0.5)to [out = 0, in = 180](7.5, 1.0)to [out = 0, in = 270](8.0, 1.5)to [out = 90, in = 0](7.5, 2.0)to [out = 180, in = 0](7.0, 1.5)to [out = 180, in = 90](6.5, 1.0);
\node[lime, above right] at (6.5, 2.0){$\TOM\left(D\right)$};
\draw[violet, fill=violet, thick] (6.5, 0.5) circle [radius = 0.05];
\draw[->, violet, thick] (6.6, 0.6) -- (8.5, 2.5);
\node[purple, below] at (8.5, 0.0){ToM moves to the UN};
\draw[red, fill=red, thick] (8.6, 2.6) rectangle (10.1, 4.1);
\node[red, below] at (10.1, 2.6){$\bTOM\left(\Ddag\right)$};
\draw[orange, fill=orange, thick] (8.6, 3.1)to [out = 270, in = 180](9.1, 2.6)to [out = 0, in = 180](9.6, 3.1)to [out = 0, in = 270](10.1, 3.6)to [out = 90, in = 0](9.6, 4.1)to [out = 180, in = 0](9.1, 3.6)to [out = 180, in = 90](8.6, 3.1);
\node[orange, above right] at (8.6, 4.1){$\TOM\left(\Ddag\right)$};
\draw[violet, fill=violet, thick] (8.6, 2.6) circle [radius = 0.05];
\end{tikzpicture} \caption{Attacker's Problem}\label{fig:diag}
\end{figure*}
In general,~\eqref{eq:eap} cannot be solved efficiently, but for reward poisoning problems with $L_{1}$ cost objective, we can relax the attacker's problem using $\iota$ strict unique Nash sets, which is a polytope described by~\eqref{eq:eisun}, and a linear outer approximation of the theory-of-mind set, a hypercube described by~\eqref{eq:eoatom}, which can be converted into a linear program and solved efficiently. We state this observation as the following proposition and depict the relationship between the sets in Figure~\ref{fig:diag}.

\begin{prop} [Reward Poisoning Linear Program] \label{prop:rplp} 
Given $\iota > 0$ and a linear $\bTOM$, the following problem is a relaxation of the attacker's reward poisoning problem and can be converted into a linear program,
\begin{align}
\displaystyle\min_{\Ddag \in \mD^{\left(R\right)}\left(D\right)} & C^{\left(1\right)}\left(D, \Ddag\right) \label{eq:erplp}
\\ \;s.t.\; & \bTOM\left(\Ddag\right) \subseteq \bNash\left(\pidag; \iota\right).\nonumber
\end{align}\end{prop}
In Figure~\ref{fig:diag}, given a dataset $D$, the general attacker's problem~\eqref{eq:eap} of moving $\TOM\left(D\right)$ {(light green)} to $\TOM\left(\Ddag\right)$ {(light red)} such that it is inside $\iNash\left(\pidag\right)$ {(light blue)} while minimizing the distance from $D$ to $\Ddag$ is often intractable. We construct a relaxed problem~\eqref{eq:erplp} of moving $\bTOM\left(D\right)$ {(green)} to $\bTOM\left(\Ddag\right)$ {(red)} such that it is inside $\bNash\left(\pidag\right)$ {(blue)}, in which all sets are polytopes and thus can be converted to a linear program for linear costs and linear theory-of-mind mappings.

In the appendix, we provide the complete linear program and show that the solution of~\eqref{eq:erplp} is feasible for~\eqref{eq:eap}. The optimality of the linear program solution depends on how close the outer approximation of the theory-of-mind set is, and in the case when the theory-of-mind set is already a hypercube, the infimum in~\eqref{eq:eap} can be achieved by taking the limit as $\iota \to  0$. 

\begin{eg} [Maximum Likelihood Centered Linear Program] \label{eg:mlclp} 
In the case $\hat{R} = R^{\text{\;MLE\;}}$ in the theory-of-mind set,~\eqref{eq:erplp} is given by,
\begin{align}
\displaystyle\min_{\rdag \in \left[-b, b\right]^{K}} & \displaystyle\sum_{k=1}^{K} \left| \rk - \rdagk \right| \label{eq:emlelp}
\\ \;s.t.\; & R^{\text{\;MLE\;}}\left(\rdag\right) \text{\;is linear in\;} \rdag \text{\;satisfying\;}~\eqref{eq:emle}\nonumber
\\ & \overline{R}\left(\rdag\right) \text{\;and\;} \underline{R}\left(\rdag\right) \text{\;satisfying\;}~\eqref{eq:eoatom}\nonumber
\\ & \hspace{2 em} \text{\;are upper and lower bounds of\;} \bTOM\left(\rdag\right)\nonumber
\\ & \left[\overline{R}\left(\rdag\right), \underline{R}\left(\rdag\right)\right] \text{\;is in\;} \bNash\left(\pidag\right) \text{\;satisfying\;}~\eqref{eq:eisun}\nonumber
\end{align}
Since $\bTOM\left(\rdag\right)$ is a hypercube and $\bNash\left(\pidag\right)$ is a polytope, the fact that the corners of the hypercube are inside the unique Nash set if and only if every element in the hypercube is in the unique Nash set implies that the constraint in~\eqref{eq:erplp} is satisfied. Technically, we only require one corner of the hypercube to be inside the unique Nash polytope, as shown in Figure~\ref{fig:diag}, and we leave the details to the proof of Proposition~\ref{prop:rplp} in the appendix. Then, because the objective and all of the constraints in~\eqref{eq:emlelp} are linear in $\rdag, \overline{R}, \underline{R}$ and $R^{\text{\;MLE\;}}$, this problem is a linear program.
\end{eg}

\section{Offline Attack on a Markov Game} 

\subsection{The Unique Nash Set (UN) of a Markov Game}
We now consider the attacker's problem for Markov games. A finite-horizon two-player zero-sum Markov game $G$ is a tuple $\left(\mS, \mA, P, R, H\right)$, where $\mS$ is the finite state space; $\mA = \mAa \times \mAb$ is the joint action space; $P = \left\{\Ph : \mS \times \mA \to  \Delta \mS\right\}_{h=1}^{H}$ is the transition function with the initial state distribution $P_{0} \in \Delta \mS$; and $R = \left\{R_{h} : \mS \times \mA \to  \left[-b ,b\right]\right\}_{h=1}^{H}$ is the mean reward function; and $H$ is the finite time horizon.

A deterministic Markovian policy $\pi = \left(\pia, \pib\right)$ is a pair of policies, where $\pii = \left\{\piih : \mS \to  \mAi\right\}_{h=1}^{H}$ for $i \in \left\{1, 2\right\}$, and $\piih\left(s\right)$ specifies the action used in period $h$ and state $s$. Again, we focus on deterministic policies, but we allow stochastic policies in which case we use the notation $\pii = \left\{\piih : \mS \to  \Delta \mAi\right\}_{h=1}^{H}$ for $i \in \left\{1, 2\right\}$, and $\piih\left(s\right)\left(\ai\right)$ represent the probability of $i$ using the action $\ai \in \mAi$ in period $h$ state $s$.

The $\text{Q}$ function is defined as, for every $h \in \left[H\right], s \in \mS, \textbf{a} \in \mA$, we write
\begin{align}
& \Qh\left(s, \textbf{a}\right) \coloneqq \Rh\left(s, \textbf{a}\right) \nonumber
\\ & \hspace{1 em} + \displaystyle\sum_{s' \in \mS} \Ph\left(s' | s, \textbf{a}\right) \displaystyle\max_{\pia \in \Delta \mAa} \displaystyle\min_{\pib \in \Delta \mAb} \Qhh\left(s', \pi\right), \label{eq:eq}
\end{align}
with the convention $\QHH\left(s, \textbf{a}\right) = 0$, and in the case $\pi$ is stochastic, we write, $\Qh\left(s, \pih\left(s\right)\right) \coloneqq$
\begin{align}
\displaystyle\sum_{a_{1} \in \mAa} \displaystyle\sum_{a_{2} \in \mAb} \piha\left(s\right)\left(a_{1}\right) \pihb\left(s\right)\left(a_{2}\right) \Qh\left(s, \left(a_{1}, a_{2}\right)\right).\nonumber
\end{align}
Given $\mS, \mA, H$, we denote the set of $\text{Q}$ functions by $\mQ = \left\{ \left\{ \Qh : \mS \times \mA \to  \mathbb{R} \right\}_{h=1}^{H} \right\}$. Technically, $\mQ$ is not the set of proper $\text{Q}$ functions of Markov games since both the reward functions and the transition functions do not have to be proper, and given $Q \in \mQ$, we may not be able to construct a Markov game that induces $Q$. This choice is made to accommodate both model-based and model-free victims who may or may not estimate the rewards and transitions explicitly from the dataset.

A stage game of a Markov game $G$ in period $h \in \left[H\right]$, state $s \in \mS$ under policy $\pi$ is a normal form game $\left(\mA, \Qh\left(s\right)\right)$, where $\mA$ is the joint action space of $G$; and $\Qh\left(s\right)$ is the mean reward function, meaning the reward from action profile $\textbf{a} \in \mA$ is $\Qh\left(s, \textbf{a}\right)$. We define Markov perfect equilibria as policies in which the action profile used in every stage game is a Nash equilibrium.

\begin{df} [Markov Perfect Equilibrium] \label{df:mpe} 
A Markov perfect equilibrium (MPE) policy $\pi$ is a policy such that $\pih\left(s\right)$ is a Nash equilibrium in the stage game $\left(\mA, \Qh\left(s\right)\right).$
\\* We define the set of all Markov perfect equilibria policies of a Markov game that induces $Q \in \mQ$ by $\fMPE\left(Q\right) = \left\{\pi : \pi \text{\;is an MPE of a Markov game with Q function\;} Q \right\}.$

\end{df}
We note that Nash equilibria for Markov games can also be defined by converting the Markov game into a single normal-form game, but we only consider Markov perfect equilibria since Nash equilibria that are not Markov perfect require coordination and commitment to policies in stage games that are not visited along equilibrium paths, which is not realistic in the MARL setting.

We define the unique Nash set for Markov games as follows.

\begin{df} [Unique Nash] \label{df:unm} 
The unique Nash set of a deterministic Markovian policy $\pi$ for a Markov game $G$ is the set of $\text{Q}$ functions such that $\pi$ is the unique Markov perfect equilibrium under policy $\pi$,
\begin{align}
\iMPE\left(\pi\right) &\coloneqq \fMPE^{-1}\left(\left\{\pi\right\}\right) = \left\{Q \in \mQ : \fMPE\left(Q\right) = \left\{\pi\right\}\right\}. \label{eq:eunm}
\end{align}\end{df}
Next, we extend the characterization of the unique Nash set for normal-form games to the Markov game setting.

\begin{thm} [Unique Nash Polytope] \label{thm:unpm} 
For any deterministic policy $\pi$,
\begin{align}
& \iMPE\left(\pi\right) = \left\{ Q \in \mQ : \pih\left(s\right) \text{\;is a strict NE of\;} \left(\mA, \Qh\left(s\right)\right),\right. \nonumber
\\ & \left.\hspace{2 em} \forall\; h \in \left[H\right], s \in \mS \right\} \nonumber
\\ &= \left\{ Q \in \mQ : \Qh\left(s, \left(\piha\left(s\right), a_{2}\right)\right) < \Qh\left(s, \pi\left(s\right)\right) \right.\nonumber
\\ &\hspace{2 em} < \Qh\left(s, \left(a_{1}, \pihb\left(s\right)\right)\right), \forall\; a_{1} \neq \piha\left(s\right),\nonumber
\\ &\left.\hspace{2em}, a_{2} \neq \pihb\left(s\right), h \in \left[H\right], s \in \mS \right\}, \label{eq:eunpm}
\end{align}\end{thm}
We show the equivalence between~\eqref{eq:eunm} and~\eqref{eq:eunpm} in the proof of Theorem~\ref{thm:unpm} in the appendix. To avoid working with strict inequalities in~\eqref{eq:eunpm}, we again define the $\iota$ strict version of the unique Nash polytope.

\begin{df} [Iota Strict Unique Nash] \label{df:isunm} 
For $\iota > 0$, the $\iota$ strict unique Nash set of a deterministic policy $\pi$ is, $\bMPE\left(\pi; \iota\right) \coloneqq$
\begin{align}
& \coloneqq \left\{ Q \in \mQ : \Qh\left(s, \left(\piha\left(s\right), a_{2}\right)\right) + \iota \leq \Qh\left(s, \pi\left(s\right)\right)\right. \nonumber
\\ &\hspace{2em} \leq \Qh\left(s, \left(a_{1}, \pihb\left(s\right)\right)\right) - \iota, \forall\; a_{1} \neq \piha\left(s\right), \nonumber
\\ &\left.\hspace{2em} a_{2} \neq \pihb\left(s\right), h \in \left[H\right], s \in \mS \right\}. \label{eq:eisunm}
\end{align}\end{df}
For every deterministic policy $\pi$ and $\iota > 0$, we have $\bMPE\left(\pi; \iota\right) \subset \iMPE\left(\pi\right)$, and the set is a polytope in $\mQ$.

\subsection{The Attacker's Theory of Mind (ToM) for Offline Multi-Agent Reinforcement Learners}
Similar to the theory-of-mind set for normal-form game learners, we define the set for Markov game learners in the $\mQ$ space. Here, $\mD$ is the set of datasets with $K$ episodes in the form $\left\{\left\{\left(\skh, \akh, \rkh\right)\right\}_{h=1}^{H}\right\}_{k=1}^{K}$ with $\skh \in \mS, \akh \in \mA$ and $\rkh \in \left[-b, b\right]$ for every $k \in \left[K\right]$, and the victims compute the Markov perfect equilibria based on the $\text{Q}$ functions estimated from such datasets.

\begin{df} [Theory of Mind] \label{df:tomm} 
Given a dataset $D \in \mD$, the theory-of-mind set $\TOM\left(D\right) \subseteq \mQ$ is the set of $\text{Q}$ functions that the victims estimate based on $D$ to compute their equilibria. In particular, if the victims learn a policy $\pi$, then $\pi \in \displaystyle\bigcup_{Q \in \TOM\left(D\right)} \fMPE\left(Q\right).$

\end{df}
\begin{eg} [Theory of Mind for Maximum Likelihood Victims] \label{eg:tommlem} 
To extend Example~\ref{eg:tommle} in the Markov game setting, we define $R^{\text{\;MLE\;}}$ the same way and $P^{\text{\;MLE\;}}$ as follows, if $\Nh\left(s, \textbf{a}\right) \coloneqq \displaystyle\sum_{k=1}^{K} \mathbb{I}_{\left\{\skh = s, \akh = \textbf{a}\right\}} > 0$,
\begin{align}
& \Rh^{\text{\;MLE\;}}\left(s, \textbf{a}|r\right) \coloneqq \dfrac{\displaystyle\sum_{k=1}^{K} \rkh \mathbb{I}_{\left\{\skh = s, \akh = \textbf{a}\right\}}}{\Nh\left(s, \textbf{a}\right)} \label{eq:emler}
\\ &\Ph^{\text{\;MLE\;}}\left(s' | s, \textbf{a}\right) \coloneqq \dfrac{\displaystyle\sum_{k=1}^{K} \mathbb{I}_{\left\{\skhh = s', \skh = s, \akh = \textbf{a}\right\}}}{\Nh\left(s, \textbf{a}\right)} \label{eq:emlep}
\\ &P_{0}^{\text{\;MLE\;}}\left(s\right) \coloneqq \dfrac{1}{K} \displaystyle\sum_{k=1}^{K} \mathbb{I}_{\left\{s^{\left(k\right)}_{1} = s\right\}} ,\nonumber
\end{align}
and if $\Nh\left(s, \textbf{a}\right) = 0$, we define $\Rh^{\text{\;MLE\;}}\left(s, \textbf{a}|r\right) \coloneqq 0$ and $\Ph^{\text{\;MLE\;}}\left(s' | s, \textbf{a}\right) \coloneqq \frac{1}{|\mS|}$.

We can construct $Q^{\text{\;MLE\;}}$ based on $R^{\text{\;MLE\;}}$ and $P^{\text{\;MLE\;}}$ according to~\eqref{eq:eq}, and since all Nash equilibria have the same value for zero-sum games, $Q^{\text{\;MLE\;}}$ is unique for every Markov perfect equilibrium of the Markov game with rewards $R^{\text{\;MLE\;}}$ and transitions $P^{\text{\;MLE\;}}$. Then we have that $\TOM\left(D\right)$ is a singleton $Q^{\text{\;MLE\;}}$.

\end{eg}
\begin{eg} [Theory of Mind for Confidence Bound Victims] \label{eg:tomcbl} 
Given a dataset $D \in \mD$, if the attacker believes the victims estimate the Markov game by estimating the rewards and transitions within some confidence region around some point estimates such as the maximum likelihood estimates, as described in~\cite{wu2022reward}, then $\TOM\left(D\right)$ would be a polytope with $\text{Q}$ functions induced by the Markov games $\left(\mS, \mA, P, R, H\right)$ with $P$ and $R$ satisfying, for every $h \in \left[H\right], s \in \mS, \textbf{a} \in \mA$,
\begin{align}
\Rh\left(s, \textbf{a}|r\right) &\in \mCRh\left(s, \textbf{a}|r\right) \label{eq:ecblr}
\\ \coloneqq &\left\{R \in \mathbb{R}: \left| R - \Rhh\left(s, \textbf{a}|r\right) \right| \leq \rhoRh\left(s, \textbf{a}\right) \right\}, \nonumber
\\ \Ph\left(s, \textbf{a}\right) &\in \mCPh\left(s, \textbf{a}\right) \label{eq:ecblp}
\\ \coloneqq &\left\{P \in \Delta \mS : \left\|P - \Phh\left(s, \textbf{a}\right)\right\|_{1} \leq \rhoPh\left(s, \textbf{a}\right)\right\}, \nonumber
\end{align}
for some point estimates $\hat{P}, \hat{R}$, and radii $\rhoR$ and $\rhoP$. We note that $\TOM\left(D\right)$ is a polytope in $\mQ$, but it has an exponential number of vertices. We can construct a tight hypercube around this polytope and call it the outer approximation of $\TOM\left(D\right)$. It contains all the $\text{Q}$ functions in the following set, for every $h \in \left[H\right], s \in \mS, \textbf{a} \in \mA$,
\begin{align}
& \Qh\left(s, \textbf{a}|r\right) \in \left[\Qlh\left(s, \textbf{a}|r\right), \Quh\left(s, \textbf{a}|r\right)\right], \label{eq:ecblq}
\\ &\Qlh\left(s, \textbf{a}|r\right) \coloneqq \displaystyle\min_{R \in \mCRh\left(s, \textbf{a}|r\right)} R \nonumber 
\\ &\hspace{1 em} + \displaystyle\min_{P \in \mCPh\left(s, \textbf{a}\right)} \displaystyle\sum_{s' \in \mS} P\left(s'\right) \displaystyle\max_{\pia \in \Delta \mAa} \displaystyle\min_{\pib \in \Delta \mAb} \Qlhh\left(s', \pi\right),\nonumber
\\ &\Quh\left(s, \textbf{a}|r\right) \coloneqq \displaystyle\max_{R \in \mCRh\left(s, \textbf{a}|r\right)} R \nonumber 
\\ &\hspace{1 em} + \displaystyle\max_{P \in \mCPh\left(s, \textbf{a}\right)} \displaystyle\sum_{s' \in \mS} P\left(s'\right) \displaystyle\max_{\pia \in \Delta \mAa} \displaystyle\min_{\pib \in \Delta \mAb} \Quhh\left(s', \pi\right).\nonumber
\end{align}\end{eg}
We omit Example~\ref{eg:tompovi} and Example~\ref{eg:tomds} for Markov games since the constructions are identical, except it is done for every stage game. As described in Example~\ref{eg:tomcbl}, we define $\Qhhh\left(s, \textbf{a}|r\right) \coloneqq \dfrac{1}{2} \left(\Quh\left(s, \textbf{a}|r\right) + \Qlh\left(s, \textbf{a}|r\right)\right)$ and $\rhoQh\left(s, \textbf{a}|r\right) \coloneqq \dfrac{1}{2} \left(\Quh\left(s, \textbf{a}|r\right) - \Qlh\left(s, \textbf{a}|r\right)\right)$, and we formally define the outer approximation of the theory-of-mind set for Markov games as follows.

\begin{df} [Outer Approximation of Theory of Mind] \label{df:oatomm} 
An outer approximation of $\TOM\left(D\right)$ is a set denoted by $\bTOM\left(D\right)$ that satisfies $\TOM\left(D\right) \subseteq \bTOM\left(D\right)$ for every $D \in \mD$, and can be written in the form,
\begin{align}
\bTOM\left(D\right) &= \left\{ Q \in \mQ : \left| \Qh\left(s, \textbf{a}\right) - \Qhhh\left(s, \textbf{a}|r\right) \right| \leq \rhoQh\left(s, \textbf{a}|r\right),\right. \nonumber
\\ & \left. \hspace{2 em} \forall\; \textbf{a} \in \mA, h \in \left[H\right], s \in \mS \right\}, \label{eq:eoatomm}
\end{align}
for some point estimate $\hat{Q}$ and radius $\rhoQ$.
\\* We call $\bTOM\left(D\right) \text{\;a\;}$ linear outer approximation if $\hat{Q}$ is linear in $\left\{\left\{\rkh\right\}_{h=1}^{H}\right\}_{k=1}^{K}$ .

\end{df}

\subsection{The Cheapest Way to Move ToM into UN for Markov Games}
In this subsection, we restate the attacker's problem for multi-agent reinforcement learners.

\begin{df} [Attacker's Problem] \label{df:apm} 
The attacker's problem with target policy $\pidag$ is,
\begin{align}
\displaystyle\inf_{\Ddag \in \mD\left(D\right)} & C\left(D, \Ddag\right) \label{eq:eapm}
\\ \;s.t.\; & \TOM\left(\Ddag\right) \subseteq \iMPE\left(\pidag\right).\nonumber
\end{align}\end{df}
For reward poisoning problems, we consider the following $L_{1}$ cost.

\begin{eg} [$L_{1}$ Cost Function] \label{eg:lcfm} 
For reward poisoning problem, where $\mD^{\left(R\right)}\left(D\right)$ is all possible datasets in the form $\Ddag = \left\{\left\{\left(\skh, \akh, \rdagkh\right)\right\}_{h=1}^{H}\right\}_{k=1}^{K}$ that are modified from $D = \left\{\left\{\left(\skh, \akh, \rkh\right)\right\}_{h=1}^{H}\right\}_{k=1}^{K}$ , we define the $L_{1}$ cost by $C^{\left(1\right)}\left(D, \Ddag\right) = \displaystyle\sum_{k=1}^{K} \displaystyle\sum_{h=1}^{H} \left| \rkh - \rdagkh \right|.$

\end{eg}
We use the same $\iota$ strictness relaxation of the unique Nash set and the linear outer approximation of the theory-of-mind set to convert~\eqref{eq:eapm} into a linear program, which can be solved efficiently. We state this observation as the following theorem.

\begin{thm} [Reward Poisoning Linear Program] \label{thm:rplpm} 
Given $\iota > 0$ and a linear $\bTOM$, the following problem is a relaxation of the attacker's reward poisoning problem and can be converted into a linear program,
\begin{align}
\displaystyle\min_{\Ddag \in \mD^{\left(R\right)}\left(D\right)} & C^{\left(1\right)}\left(D, \Ddag\right) \label{eq:erplpm}
\\ \;s.t.\; & \bTOM\left(\Ddag\right) \subseteq \bNash\left(\pidag; \iota\right).\nonumber
\end{align}\end{thm}
\begin{eg} [Maximum Likelihood Centered Linear Program] \label{eg:mlclpm} 
In the case $\hat{R} = R^{\text{\;MLE\;}}$ and $\hat{P} = P^{\text{\;MLE\;}}$, and we construct $\bTOM\left(D\right)$ as described in Example~\ref{eg:tomcbl},~\eqref{eq:erplpm} can be converted into a linear program even without explicitly constructing the $\bTOM\left(D\right)$ set. We provide an intuition here and the formal construction in the proof of Theorem~\ref{thm:rplpm},
\begin{align}
\displaystyle\min_{\rdag \in \left[-b, b\right]^{K}} & \displaystyle\sum_{k=1}^{K} \displaystyle\sum_{h=1}^{H} \left| \rkh - \rdagkh \right| \label{eq:emlclpm}
\\ \;s.t.\; & R^{\text{\;MLE\;}}\left(\rdag\right) \text{\;is linear in\;} \rdag \text{\;satisfying\;}~\eqref{eq:emler}\nonumber
\\ & P^{\text{\;MLE\;}} \text{\;is independent of\;} \rdag \text{\;satisfying\;}~\eqref{eq:emlep}\nonumber
\\ & Q^{\text{\;MLE\;}}\left(\rdag\right) \text{\;satisfying\;}~\eqref{eq:eq} \nonumber
\\ & \hspace{2 em} \text{\;is linear in\;} R^{\text{\;MLE\;}}\left(\rdag\right) \text{\;thus\;} \rdag \nonumber
\\ & \overline{Q}\left(\rdag\right) \text{\;and\;} \underline{Q}\left(\rdag\right) \text{\;satisfying\;}~\eqref{eq:ecblq} \nonumber
\\ & \hspace{2 em} \text{\;are upper and lower bounds of\;} \bTOM\left(\rdag\right) \nonumber
\\ & \left[\overline{Q}\left(\rdag\right), \underline{Q}\left(\rdag\right)\right] \text{\;is in\;} \bNash\left(\pidag\right) \text{\;satisfying\;}~\eqref{eq:eisunm}\nonumber
\end{align}
We move the hypercube $\bTOM\left(\rdag\right)$ into the polytope $\bNash\left(\pidag\right)$ by moving one of the corners into the polytope. Note that if $\overline{Q}$ and $\underline{Q}$ are not constructed directly as linear functions of $\rdag$, and are computed by~\eqref{eq:ecblq}, then these constraints are not linear in $\rdag$. We avoid this problem by using the dual linear program of~\eqref{eq:ecblq}. We present the details in the appendix in the proof of Theorem~\ref{thm:rplpm}. All other constraints are linear in $\rdag$, and as a result,~\eqref{eq:emlclpm} is a linear program.

\end{eg}
In the end, we present a sufficient but not necessary condition for the feasibility of~\eqref{eq:erplpm} and~\eqref{eq:eapm}. This condition applies directly to normal-form games with $H = 1$.

\begin{thm} [Reward Poisoning Linear Program Feasibility] \label{thm:rplpf} 
For $\iota > 0$, $\TOM\left(D\right)$ with $\hat{Q} = Q^{\text{\;MLE\;}}$, and $\Nh(s, \textbf{a}) > 0$ for every $h \in \left[H\right], s \in \mS, \textbf{a} \in \mA$ where either $a_1 = \pidagha\left(s\right)$ or $a_2 = \pidaghb\left(s\right)$, the attacker's reward poisoning problem is feasible if for every $h \in \left[H\right], s \in \mS, \textbf{a} \in \mA$,
\begin{align}
\rhoRh\left(s, \textbf{a}\right) &\leq \dfrac{b - \iota}{4 H}. \label{eq:erplpf}
\end{align}\end{thm}

\begin{table}
\small
\begin{minipage}{.49\linewidth}
\begin{center}\begin{tabular}{|c|c|c|c|c|}
\hline
 $\mAa \setminus \mAb$ &$1^{\dagger}$ &$2$ &$3$\\ \hline
$1^{\dagger}$ &$0$ &$b$ &$b$\\ \hline
$2$ &$-b$ &-&-\\ \hline
$3$ &$-b$ &-&-\\ \hline
\end{tabular}\end{center}
\caption{A Feasible Attack} \label{tab:feas}
\end{minipage}
\begin{minipage}{.49\linewidth}
\begin{center}\begin{tabular}{|c|c|c|c|}
\hline
 $\mA$ &$H$ &$T$\\ \hline
$H$ &$U\left[0, 1\right]$ &$U\left[-1, 0\right]$\\ \hline
$T$ &$U\left[-1, 0\right]$ &$U\left[0, 1\right]$\\ \hline
\end{tabular}\end{center}
\caption{The original dataset generation distributions} \label{tab:smp}
\end{minipage}
\end{table}

To construct a feasible attack under~\eqref{eq:erplpf}, we use the poisoned rewards similar to the one shown in Table~\ref{tab:feas}, which is an example where each agent has three actions and the target action profile being action $\left(1, 1\right)$. With this $\rdag$, the maximum likelihood estimate of the game has a unique Nash equilibrium $\pidagh\left(s\right)$ with a value of $0$ in every stage $\left(h, s\right)$. Furthermore, if either the radius of rewards or the radius of $\text{Q}$ functions for the theory-of-mind set is less than $\frac{b - \iota}{4 H}$, we can show inductively that $\pidagh\left(s\right)$ remains the unique Nash equilibrium in every stage $\left(h, s\right)$, thus showing that every $\text{Q}$ function in the theory-of-mind set is also in the unique Nash set, which means the attack is feasible. The complete proof is in the appendix.

\section{Experiments} 

\subsection{Rock Paper Scissors}
We start with a simple toy dataset for the Rock Paper Scissors (RPS) game, shown in Table~\ref{tab:rps} with partial coverage, where each entry appears once in the dataset, and the target action profile is $\pi^{\dagger} = \left(R, R\right)$, leading to a tie.
\newline
\begin{table}[h]
\small
\begin{minipage}{.38\linewidth}
\begin{center} \begin{tabular}{|c|c|c|c|c|}
\hline
  &$R$ &$P$ &$S$\\ \hline
$R$ &$0$ &$-1$ &$1$\\ \hline
$P$ &$1$ &$0$ &$-1$\\ \hline
$S$ &$-1$ &$1$ &$0$\\ \hline
\end{tabular} \end{center} \caption{RPS Game} \label{tab:rps}
\end{minipage}
\begin{minipage}{.28\linewidth}
\begin{center} \begin{tabular}{|c|c|c|c|c|}
\hline
$R$ &$P$ &$S$\\ \hline
$0$ &$-1$ &$1$\\ \hline
$1$ &- &-\\ \hline
$-1$ &- &-\\ \hline
\end{tabular} \end{center} \caption{Original} \label{tab:rpso}
\end{minipage}
\begin{minipage}{.3\linewidth}
\begin{center} \begin{tabular}{|c|c|c|c|c|}
\hline
$R$ &$P$ &$S$\\ \hline
$0$ &$\iota$ &$1$\\ \hline
$-\iota$ &- &-\\ \hline
$-1$ &- &-\\ \hline
\end{tabular} \end{center} \caption{Poisoned} \label{tab:rpsp}
\end{minipage}
\end{table}

Given the original dataset with $5$ entries described in Table~\ref{tab:rpso}, our algorithm with $\rho=0$ and $\iota=0.01$ leads to the poisoned dataset described in Table~\ref{tab:rpsp}. The attack cost is $2.02$, whereas the attack cost from the feasible attack described in Table~\ref{tab:feas} with $b=1$ is $4$. In addition, note that given the partial coverage, the attack described in ~\cite{wu2022reward} is not feasible due to their full coverage requirement.

\subsection{Stochastic Matching Penny}
We follow up with the matching penny game, which is also the penalty kick game in soccer, and the rewards are usually estimated by random data points. We generate the datasets randomly with Uniform distributions summarized in Table~\ref{tab:smp}. The attacker would like to install a target action profile of $\left(H, H\right)$, and in the context of the penalty kick game, the attacker's motivation might be to increase or decrease the total number of goals.

We summarize the before-vs-after box plots in Figure~\ref{fig:bbx} for the $n = 100$ case. The cost comparison of our attack, the feasible attack in Table~\ref{tab:feas} with $b=1$, and the Dominant Strategy Equilibrium (DSE) attack in ~\cite{wu2022reward}, is given in Table~\ref{tab:tbx}.

\begin{figure}
\centering
\begin{subfigure}{.2\textwidth}
\centering
\begin{tikzpicture} [scale = 0.75]
\begin{axis}[
    boxplot/draw direction=y,
	xtick={1,2,3,4},
	xticklabels={HH,HT,TH,TT},
]
    \addplot+ [boxplot prepared={
        lower whisker=-1.54, lower quartile=-0.75,
        median=-0.48, upper quartile=-0.23,
        upper whisker=0.56},
    ] coordinates {};
    \addplot+ [boxplot prepared={
        lower whisker=-0.55, lower quartile=0.22,
        median=0.47, upper quartile=0.74,
        upper whisker=1.51},
    ] coordinates {};
	\addplot+ [boxplot prepared={
        lower whisker=-0.62, lower quartile=0.20,
        median=0.47, upper quartile=0.75,
        upper whisker=1.58},
    ] coordinates {};
	\addplot+ [boxplot prepared={
        lower whisker=-1.28, lower quartile=-0.69,
        median=-0.49, upper quartile=-0.29,
        upper whisker=0.30},
    ] coordinates {};
\end{axis}
\end{tikzpicture}
\caption{Original} \label{fig:bbx}
\end{subfigure}
\begin{subfigure}{.2\textwidth}
\centering
\begin{tikzpicture} [scale = 0.75]
\begin{axis}[
    boxplot/draw direction=y,
	xtick={1,2,3,4},
	xticklabels={HH,HT,TH,TT},
]
    \addplot+ [boxplot prepared={
        lower whisker=-1.54, lower quartile=-0.75,
        median=-0.48, upper quartile=-0.23,
        upper whisker=0.56},
    ] coordinates {};
    \addplot+ [boxplot prepared={
        lower whisker=-3.36, lower quartile=-1,
        median=-1, upper quartile=0.57,
        upper whisker=2.93},
    ] coordinates {};
	\addplot+ [boxplot prepared={
        lower whisker=-0.62, lower quartile=0.20,
        median=0.47, upper quartile=0.75,
        upper whisker=1.58},
    ] coordinates {};
	\addplot+ [boxplot prepared={
        lower whisker=-1.36, lower quartile=-0.68,
        median=-0.45, upper quartile=-0.22,
        upper whisker=0.46},
    ] coordinates {};
\end{axis}
\end{tikzpicture}
\caption{Poisoned} \label{fig:abx}
\end{subfigure}
\caption{Distribution of rewards} \label{fig:cbx}
\end{figure}

\begin{table}
\begin{center} \begin{tabular}{|c|c|c|c|c|}
\hline
 Average costs &$n = 1$ &$n = 10$ &$n = 100$\\ \hline
Our attack &1.06 &9.09 &99.47\\ \hline
Feasible attack &2.12 &16.08 &250.46\\ \hline
DSE attack &2.06 &18.31 &198.38\\ \hline
\end{tabular} \end{center}
\caption{Cost comparison between different attacks} \label{tab:tbx}
\end{table}

\section{Discussions} 
We discuss a few extensions. Faking a unique mixed strategy Nash equilibrium is in general impossible due to the sensitivity of mixing probabilities from small perturbations of the reward function, and as long as the theory-of-mind set has non-zero volume, it is impossible to install a mixed strategy profile (or stochastic policy for Markov games) as the unique equilibrium. Faking a unique optimal policy for single-agent reinforcement learners can be easily adapted from our linear program~\eqref{eq:erplpm}. Faking a unique coarse correlated equilibrium in every stage game is equivalent to our problem as well since for a two-player zero-sum game, a policy is the unique Markov perfect coarse correlated equilibrium if and only if it is the unique Markov perfect Nash equilibrium.

\section{Acknowledgments}
This project is supported in part by NSF grants 1545481, 1704117, 1836978, 1955997, 2023239, 2041428, 2202457, ARO MURI W911NF2110317, and AF CoE FA9550-18-1-0166, and we thank Yudong Chen for his useful comments and discussions.

\bibliography{aaai24}




\end{document}